\documentclass[12pt,preprint]{aastex}
\newcommand{\jcd}{Christensen-Dalsgaard}
\newcommand{\Rs}{R_{\odot}}

\shorttitle{How do $f$-mode Frequencies Change with Solar Radius?}
\shortauthors{Chatterjee \& Antia}

\begin{document}

\title{How do $f$-mode Frequencies Change with Solar Radius?}

\author{Piyali Chatterjee and H. M. Antia}
\affil{Department of Astronomy and Astrophysics, Tata Institute of Fundamental Research, Colaba, Mumbai 400005, India}

\email{piyalic@tifr.res.in, antia@tifr.res.in}

\begin{abstract}
We test the relation between relative $f$-mode frequency variation ($\delta \nu/ \nu$) and Lagrangian perturbation in the solar radius ($\delta r/ r$) obtained by Dziembowski \& Goode (2004) using several pairs of solar models and show that it doesn't hold true for any of the model pairs we have used.
We attempt to derive a better approximation for the kernel linking the relative frequency changes and the solar radius variation in the subsurface layers.
\end{abstract}

\keywords{Sun: oscillations -- Sun: helioseismology}

\section{Introduction}
The solar $f$-modes which are essentially surface gravity modes,
provide a diagnostic of flows and magnetic fields present in the near surface region (Murawski \& Roberts 1993; Rosenthal \& Gough 1994; Rosenthal \& Christensen-Dalsgaard 1995; Sofia et al.~2005).
The $f$-mode frequencies can also provide an accurate measure of solar radius (Schou et al.~1997; Antia 1998).
The dependence of solar radius on the 11-year activity cycle is still a matter of controversy.
Different measurements of solar radius have given conflicting results
about its temporal variations (e.g., Laclare et al.~1996; No\"el, 2004; Kuhn et al.~2004; Chapman et al.~2008; Djafer et al.~2008).
Using the energy budget of the Sun, it is easy to see that the radius variation, if any, would be localized only in the near surface regions which are well sampled by $f$-modes.
Hence, we can expect the $f$-mode frequencies to reflect these variations in the solar
radius.
Dziembowski et al.~(2001) and Dziembowski \& Goode (2004) (henceforth DG04)
have obtained a relation between the $f$-mode frequency variations
and radius variations in the subsurface layers.
This relation was later used by Lefebvre \& Kosovichev (2005, 2007) to show that helioseismic radius varies in anti-phase with solar activity
in the outer region of the Sun but there is a change in behavior in deeper layers.

However to the best of our knowledge this equation (cf., Eq.~(1)) has not
been verified. In this work, we attempt to test this equation using a few
pairs of solar models to check if the actual frequency differences match
the values computed from Eq.~(1).

The paper is organized as follows: In \S2 we test the relation between $\delta \nu_l/\nu_l$ and Lagrangian radius perturbation ($\delta r/r$) as derived by DG04 using five different solar models. While in \S3 we derive a different kernel to describe the above relation and test it with the ten model pairs. \S4 summarizes the main conclusions from this study.

\section{Testing frequency-radius relation with Solar models}

	DG04 obtained the following relation between the relative $f$-mode 
frequency change and the Lagrangian radius variation: 
\begin{equation}
\frac{\delta \nu_l}{\nu_l}= -\frac{3 l}{2\omega_l^2 I_l}\int dI_l \frac{g}{r}\frac{\delta r}{r}, 
\label{eq:dg}
\end{equation}
where $g$ is the acceleration due to gravity,
$I_l$ is the mode inertia of $f$-mode with degree $l$ and $\omega_l = 2\pi \nu_l$ is the angular frequency. If we take two solar models 
having the same radius $\Rs$ but differing in structure through differences
in input physics, like the equation of state or the
treatment of convective flux, then in the subsurface layers the radius
enclosing the same mass will in general not be the same.
We can use
such models to test Eq.~(\ref{eq:dg}) by comparing the frequency differences
to that obtained from the integral on the right hand side.
In that case $\delta \nu_l$ is the 
frequency difference of the linear adiabatic oscillations of degree $l$ and order 0 ($f$-modes) 
between the models and $\delta r$ is the difference in radius at constant mass for the same pair of 
models. We calculate the eigenfrequencies $\nu_l$ and the eigenfunctions
${\mathbf \xi}_l$ for all the $f$-modes with $40\le l\le 1000$ for all the
models listed in Table~1 using a stellar adiabatic pulsation code. Solar f-modes have been recently observed up to a degree of 900 (Rabello-Soares et al.~2008).
In these models the mesh spacing near the surface was reduced till there was
no difference in the results to ensure that numerical accuracy is sufficient
to describe the high degree $f$-modes.
All these models have a solar radius of 695780 km and
use OPAL opacity tables (Iglesias \& Rogers 1996).
The heavy element abundance $Z$ at the surface is 0.0179 in all the models except OPALOWZ for which $Z = 0.0127$.
We have also used models with different equations of states (EOS), i.e., the
OPAL (Rogers et al.~1996; Rogers \& Nayfonov 2002),
MHD (D\"appen et al.~1988; Hummer \& Mihalas, 1988; Mihalas et al.~1988)
and CEFF (Eggleton et al.~1973; \jcd\ \& D\"appen 1992; Guenther et al.~1992).
For calculating convective flux, we use either the Mixing Length Theory (MLT)
or the formulation due to Canuto \& Mazzitelli (1991) (CM).
We then compare the left hand side and the right hand side of Eq.~(\ref{eq:dg}) for all the ten possible pairs
of models.
Fig.~(1b) shows the result for the pair OPALCM -- OPALMLT. The models OPALCM and OPALMLT are very similar to each other in that the relative frequency difference is of the order of $10^{-5}$ compared to all other pairs where the same difference is $\sim 10^{-4}$. Even then it may be noted from Fig.~(1b) that Eq.~(\ref{eq:dg}) is a poor approximation.

The inconsistency in Eq.~(\ref{eq:dg}) is very prominent in Fig.~(2b) where we have compared models MHD and OPALOWZ (see Table~1).
For this pair $\delta \nu/\nu$ changes sign near $l = 80$ where as the integral on the right hand side of Eq.~(\ref{eq:dg}) doesn't. This is so because $\delta r/r$ doesn't change sign in the subsurface layers and hence the integral is always positive (see Fig.~(2a)).

Even though we have shown only two of the ten cases for which we have tested
Eq.~(\ref{eq:dg}), we found similar disagreement for the other eight pairs.
Hence it is unlikely, that this equation is valid for the Sun and
there is considerable doubt about the validity of the results obtained
by inverting Eq.~(\ref{eq:dg}) to calculate the solar seismic radius
variation with time.

\section{Deriving the Kernels for $\delta r$}
The equations governing linear adiabatic stellar oscillations are given by (cf.,~Unno et al.~1989)
\begin{eqnarray}
\nonumber
-\omega^2\rho {\bf \xi} = \nabla(c^2\rho\nabla\cdot{\bf \xi}+\nabla p\cdot{\bf \xi})-\\
{\bf g} \nabla\cdot(\rho {\bf \xi})-G\rho\nabla \left(\int_{V}\frac{\nabla\cdot(\rho \xi)d^3 r'}{|r-r'|}\right),
\label{eq:unno}
\end{eqnarray}
where $\rho$, $p$, $c$ and $\bf g$ are the density, pressure, sound speed and acceleration due to gravity in the stellar model.
It was precisely this eigenvalue problem which was solved by the adiabatic pulsation code in \S2.

The variational formulation of this equation (Chandrasekhar 1964)
has been used to find relation between frequency variations and structure
variations (e.g., Dziembowski et al.~1990; Antia \& Basu 1994; henceforth AB94). These relations
have been obtained by linearizing the variational formulation about a
reference solar model and are extensively used for structure inversions. 
These relations have also been tested by using a pair of solar models
in a similar manner to our test in \S2.

Since these relations are obtained by linearizing the differences, they are
expected to be valid between pairs of models where the differences in
structure variables like sound speed and density are small.
The relative frequency difference, $\delta\nu/\nu$, between two
models can be written in terms
of four integrals, $I_1,I_2,I_3,I_4$ as defined in Eq.~(8) of AB94. For the sake of completeness we provide the expressions of $I_1$ and $I_3$ here.
\begin{eqnarray}
I_1&=& -\int_{0}^{\Rs}\rho(\nabla\cdot{\mathbf \xi})^2 \delta c^2 r^2\;dr,
\label{eq:I1}\\
I_3 &=&\int_{0}^{\Rs}\rho c^2\xi_r \nabla\cdot{\mathbf \xi}  \frac{d}{dr}\left(\frac{\delta \rho}{\rho}\right) r^2\;dr.
\label{eq:I3}
\end{eqnarray}
It appears that for the $f$-modes in models listed in Table~1
the major contribution comes from the integral $I_3$ involving $\delta \rho/\rho$ and the second major contribution comes from $I_1$ involving $\delta c^2/c^2$.
Fig.~(3) shows $\delta \nu/\nu$, $-(I_1+I_3)/2\omega^2$ and $-I_3/2\omega^2$ as a function of $l$ for model combination OPALCM -- OPALMLT. The contributions from $I_2$ and $I_4$ are negligible supporting the use of Cowling's approximation by DG04. 

As noted above $I_3$ makes the dominant contribution to $\delta \nu/\nu$.
Let $\delta_r$ denote the variation at fixed radius in contrast to $\delta_m$ which denotes the change at constant mass (the Lagrangian variation).
The integral $I_3$ includes the derivative of $\delta_r \rho/\rho$ with respect to radial distance. We now attempt to express  $\delta_r \rho/\rho$ in terms of $\delta_m r$. Using Taylor's theorem we can write:
\begin{equation}
\frac{\delta_m \rho}{\rho} = \frac{\delta_r \rho}{\rho} -\frac{\delta_m r}{H_{\rho}},
\label{eq:den1}
\end{equation}
where $H_{\rho}$ is the density scale height. Also conservation of mass gives the following relation:
\begin{equation}
\frac{\delta_m \rho}{\rho} = -\frac{1}{r^2}\frac{d}{dr}\left(r^2 \delta_m r\right).
\label{eq:den2}
\end{equation}
Combining Eq.~(\ref{eq:den1}) with Eq.~(\ref{eq:den2}) we get:
\begin{equation}
\frac{\delta_r \rho}{\rho} =  -\frac{1}{r^2}\frac{d}{dr}\left(r^2 \delta_m r\right) +\frac{\delta_m r}{H_{\rho}}.
\label{eq:den3}
\end{equation}
Using Eq.~(\ref{eq:den3}) and an integration by parts on integral $I_3$ in Eq.~(\ref{eq:I3}) we finally have:
\begin{equation}
\tilde{I_3} = - \int_{0}^{\Rs} \left \{\frac{K_3'}{H_{\rho}}- \frac{2 K_3'}{r} + K_3''\right \}\delta_m r dr.
\label{eq:tI3}
\end{equation}
Here $K_3 = r^2\rho c^2 \xi_r \nabla\cdot{\mathbf\xi}$ and the primes denote the derivative with respect to $r$.

Transforming the kernel $K_1$ of $\delta c^2/c^2$ in the integral $I_1$ into a kernel of $\delta_m r$ is more algebraically involved. Moreover we make the assumption that $c^2 = (\partial p/ \partial \rho)_S = p'/\rho'$ for the subsurface region.
This relation may not be true very near the surface where the stratification
is not adiabatic. Using hydrostatic balance we can write:
\begin{equation}
\frac{\delta_r c^2}{c^2} = H_{\rho}\frac{d}{dr}\left(\frac{\delta_r \rho}{\rho}\right) + 
\frac{\delta_r g}{g}.
\label{eq:c2}
\end{equation}
Since $g \propto m/r^2$, the above equation may be transformed in terms of $\delta_m r$ using
\begin{equation}
\frac{\delta_r g}{g}= -\frac{4\pi \rho r^3}{m}\frac{\delta_m r}{r}.
\label{eq:g}
\end{equation}
However we neglect the term $\delta_r g/g$ as it is a small fraction of $\delta_m r/r$ since $4\pi \rho r^3/m$ varies from $10^{-3}$ at 0.97$\Rs$ to $10^{-6}$ near the surface. Finally a couple of integrations by parts are required on $I_1$ due to the presence of the derivative in the first term of Eq.~(\ref{eq:c2}) to write it completely in terms of $\delta_m r$ as
\begin{eqnarray}
\tilde{I_1} = - \int_{0}^{\Rs} \left \{\frac{K_1'}{H_{\rho}}- \frac{2 K_1'}{r}
 + K_1''\right \}\delta_m r \;dr,
 \label{eq:tI1}
\end{eqnarray}
where $K_1 = -\rho (\nabla\cdot{\mathbf \xi})^2 c^2 H_{\rho} r^2$. A word of caution regarding using the kernel in 
$\tilde{I_1}$ here is necessary. These kernels can have errors very near to the surface, $r> 0.998\Rs$ because in the outer part of the convection zone the temperature gradient is
not close to adiabatic value, that is required for the relation
$c^2=p'/\rho'$ to be valid.
We have compared the terms $-\tilde{I_3}/2\omega^2$ and $-(\tilde{I_1}+\tilde{I_3})/2\omega^2$ with $\delta \nu/\nu$ in Fig.~(1b) for model pair OPALCM -- OPALMLT and in Fig.~(2b) for the pair MHD -- OPALOWZ. 
It is evident that kernels given in Eq.~(\ref{eq:tI3}) and Eq.~(\ref{eq:tI1}) are in much better agreement with $\delta\nu/\nu$ than that given 
by DG04 (our Eq.~(\ref{eq:dg})). From Fig.~(2b) it can be seen that $\delta \nu/\nu$
changes sign at $l \approx 80$. This is because the slope of $\delta \rho/\rho$ changes sign around $r = 0.98 \Rs$ (dashed line in Fig.~(2a)). There also exists sign change in the slope beyond $r = 0.998\Rs$, which is manifested as a sign change in $\delta \nu/\nu$ at $l \sim 450$. In contrast $\delta_m r/r$ (solid line in Fig.~(2a)) decreases monotonically and hence the sign change in $\delta \nu/\nu$ cannot be captured by the right hand side of Eq.~(1).
In fact no equation of this form with a kernel that doesn't change sign can
account for the differences in this case.
\section{Conclusions}
We have shown that the relation connecting the variation in relative frequency and the Lagrangian radius variation obtained by Dziembowski \& Goode (2004) does not hold good for any of the ten model combinations considered by us.
On the other hand, the kernel derived by us gives results in which the discrepancies are much smaller for all
the model pairs considered in this work.
It may be noted that all the models considered by us have the same solar radius even though they differ in their subsurface variation in radius at constant 
mass ($\delta_m r$). In fact Sofia et al.~(2005), and Lefebvre \& Kosovichev (2005) point out that the Solar radius change with solar activity must be non-homologous in the subsurface layers. However we would like to point out that in that case the idea of solar radius variation and its observational evidence becomes highly ambiguous.
It is not clear if the use of such radius variation would give any insight
into the structural variations in the solar interior. Moreover Eq.~(\ref{eq:dg}) as
well as the relation derived in this work, does not include the effect of
magnetic field. Thus the effect of magnetic field can be mis-interpreted
as due to radius change by the use of these equations. Of course, the
presence of magnetic field will induce structural variations, which may be
accounted for by these relations, but in addition there will be contribution
to frequency variation due to the direct effect of magnetic field, through
the Lorentz force, which cannot be accounted for by such relations.

%\acknowledgments

\clearpage

\begin{table}
\begin{center}
\caption{Overview of the solar models used}
\vspace{0.5cm}
\begin{tabular}{crrrr}
\tableline\tableline
&Model & EOS & Convection & Z \\
\tableline
1 & OPALCM & OPAL & CM & 0.0179\\ 
2 & OPALMLT & OPAL & MLT & 0.0179\\ 
3 & OPALOWZ & OPAL & CM & 0.0127\\ 
4 & MHD &MHD  &  CM & 0.0179\\ 
5 & CEFF &CEFF & CM & 0.0179\\ 
\tableline
\end{tabular}
\end{center}
\end{table}

\clearpage
\begin{figure}[h!]
\centering{\includegraphics[width=0.8\textwidth]{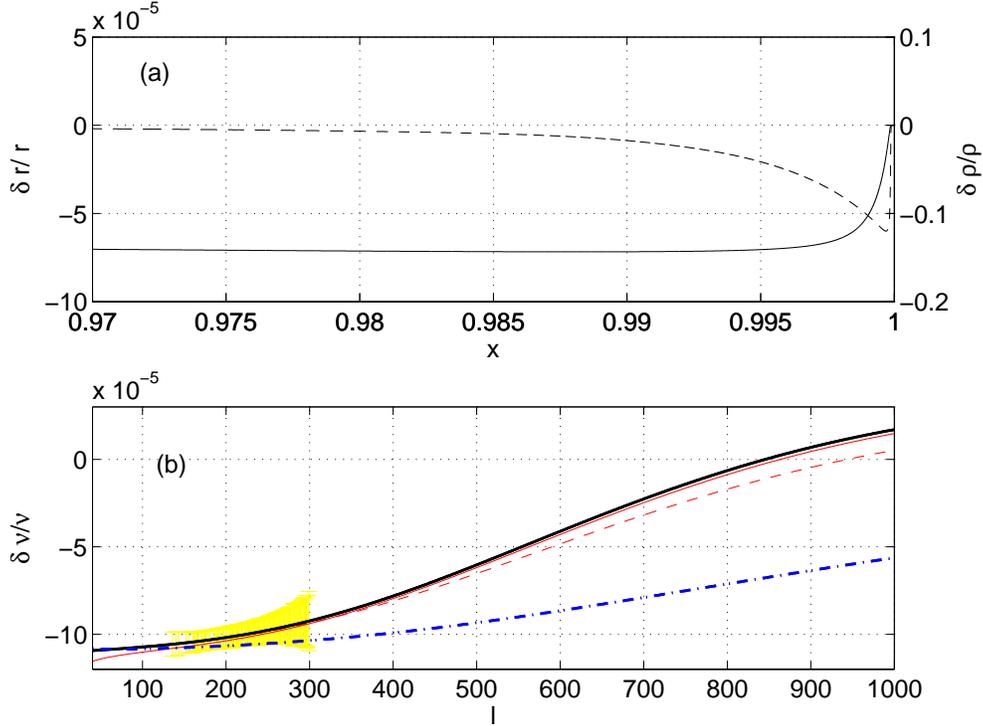}}
\caption{Comparison between models OPALCM and OPALMLT. (a) $\delta r/r$ (solid line) at constant mass as a function of $x = r/\Rs$ for the same pair. The dashed line gives $\delta \rho/\rho$ at constant radius as a function of $x$. (b) $\delta \nu/\nu$ (thick solid),
right hand side of Eq.~(1) (dashed dotted), $-(\tilde{I_1}+\tilde{I_3})/2\omega^2$ (dashed), $-\tilde{I_3}/2\omega^2$ (thin solid) as a function of $l$ ({\em see text for definitions}). The error bars denote the errors for observed $f$-mode frequencies (Schou 1999). Note that the result deteriorates by adding $\tilde{I_1}$ to $\tilde{I_3}$. As mentioned in \S3, $\tilde{I_1}$ is not a good approximation to $I_1$ in the outermost layers where this pair of models have
maximum differences. }
\end{figure}
\begin{figure}[h!]
\centering{\includegraphics[width=0.8\textwidth]{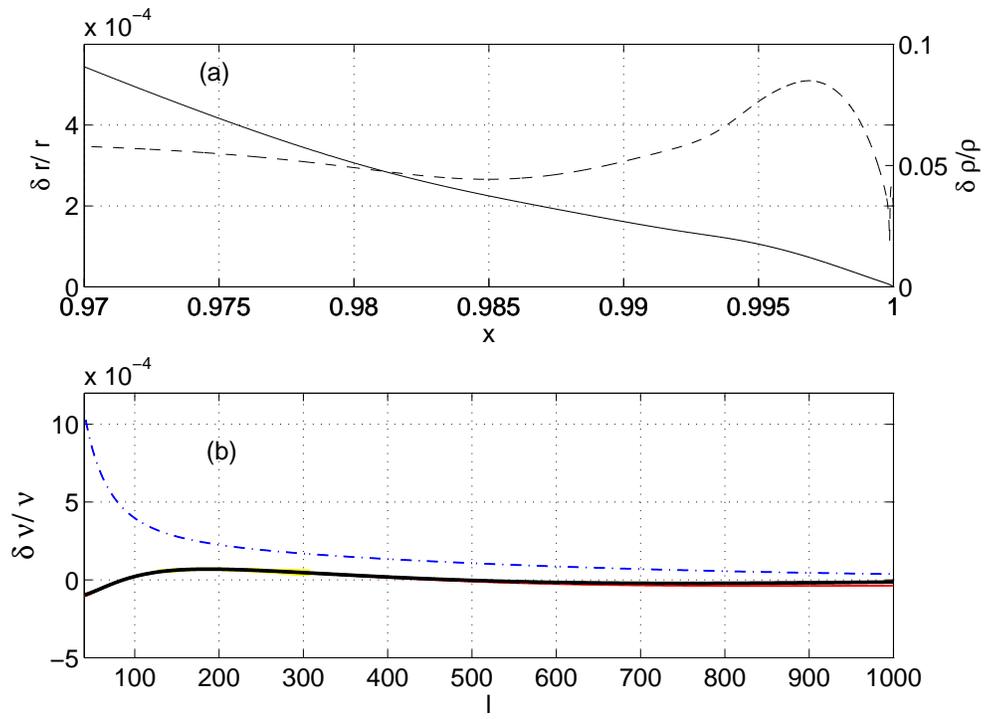}}
\caption{Same as Fig.~1 but for models MHD and  OPALOWZ}
\end{figure}
\clearpage
\begin{figure}[h!]
\centering{\includegraphics[width=0.7\textwidth]{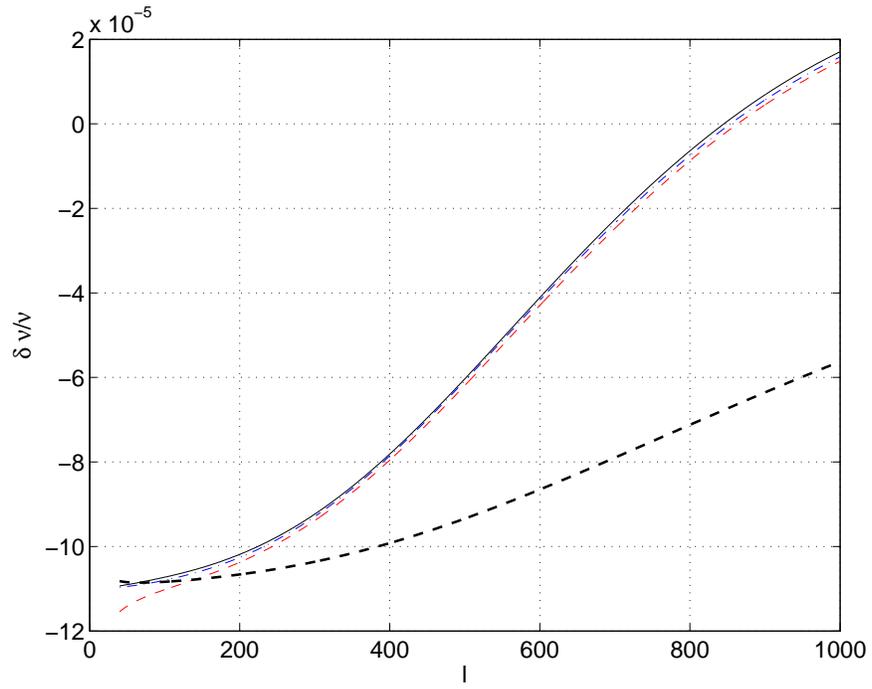}}
\caption{$\delta \nu/\nu$ (solid) for OPALCM -- OPALMLT,
summation of integrals (from AB94) $-(I_1+I_3)/2\omega^2$ (dashed dotted), $-I_3/2\omega^2$ (dashed) as a function of $l$. Compare this with right hand side of Eq.~(1) (thick dashed) as a function of $l$.}
\end{figure}

\end{document}